\documentclass{iaus}
\usepackage{graphicx}

\title[Evolution of the First Stars] 
{Evolution of the First Stars:  CNO Yields and the C-rich Extremely Metal Poor Stars}

\author[Meynet et al.]   
{Georges Meynet, Sylvia Ekstr\"{o}m, Andr\'e Maeder}%

\affiliation{Geneva Observatory, CH--1290 Sauverny, Switzerland
\break email: georges.meynet@obs.unige.ch 
\break email: sylvia.ekstrom@obs.unige.ch 
\break email:andre.maeder@obs.unige.ch 
}

\pubyear{2005}
\volume{228}  
\pagerange{1--7}
\date{?? and in revised form ??}
\jname{From Lithium to Uranium: Elemental Tracers of Early Cosmic Evolution}
\editors{V. Hill, P. Fran\c{c}ois \& F. Primas, eds.}
\begin{document}

\maketitle

\begin{abstract}
Rotating massive stars at $Z=10^{-8}$ and $10^{-5}$ 
lose a great part of their initial mass through stellar winds.
The chemical composition of the rotationally enhanced winds of very low $Z$ stars is very peculiar. 
The winds show large CNO enhancements  by factors of $10^3$ to $10^7$, together with large excesses of $^{13}$C
and $^{17}$O  and moderate amounts of Na and Al. The excesses of primary N are particularly striking.
When these ejecta from the rotationally
enhanced winds are diluted with the supernova ejecta from the corresponding CO cores, we find [C/Fe], [N/Fe],[O/Fe]
abundance ratios very similar to those observed in the C--rich extremely metal poor stars (CEMP).
We show that rotating AGB stars and rotating massive stars have about the same effects on the CNO enhancements.
Abundances of s-process elements and the $^{12}$C/$^{13}$C ratio could help us to distinguish between contributions
from AGB and massive stars. 
On the whole, we emphasize the dominant effects of rotation for the chemical yields 
of extremely metal poor stars.
\keywords{stars: abundances, stars: rotation, stars: mass loss, stars: AGB and post-AGB}
\end{abstract}

\firstsection 
              
\section{Some puzzling observational facts}

Many intriguing and fascinating results were shown during this conference, which
certainly require some revision of our classical understanding of the evolution
of the first stellar generations in the Universe. Let us briefly recall a few of them here:
\begin{itemize}

\item The measured abundances of many elements at very low metallicity present
a very small scatter (Cayrel et al.~\cite{cayr04}).
This might indicate that already at this low metallicity, stars are formed from 
a well mixed reservoir composed of ejecta from stars of different initial masses.

\item At least down to a metallicity of [Fe/H] equal to -4, there is no sign of enrichments
by pair instability supernovae (see the contribution by Heger in this volume).

\item The N/O ratios observed at the surface of halo stars by Israelian et al.~(\cite{Is04}) and
Spite et al. (\cite{Sp05})
indicate that important amounts of primary nitrogen should be produced by 
very metal-poor massive stars
(Chiappini et al.~\cite{Ch05}).

\item Below [O/H] $< -2$ the C/O ratio presents an upturn indicating
either that very metal poor stars produce more carbon, or less oxygen or both 
(Akerman et al.~\cite{Ak04}; Spite et al.~\cite{Sp05})

\item If most stars, at a given [Fe/H], present a great
homogeneity in composition, a small group, comprising about 20 - 25\% of the stars
with [Fe/H] below -2.5, show very large enrichments in carbon with a great scatter,
these are the C-rich Extremely Metal Poor stars (CEMP).

\item The surface abundances of stars in globular clusters show an anti correlation
between the abundances of sodium and oxygen, those of magnesium and aluminum (see the contributions
by Gratton and Charbonnel in this volume). Halo stars in the field do not show such
anti correlations.

\item The zero age main sequence in the very massive globular cluster $\omega$ Cen can
be decomposed in a blue and a red component. The blue component is paradoxically more metal
rich by about a factor two with respect to the red one. 
The only way, presently, to understand such a behavior, is to
suppose that the stars of the blue component are very helium-rich
(see the contributions by Gratton and Maeder et al. in this volume).

\end{itemize} 

Many more results could be added to the above list. Such a large amount of unexplained
features points toward the need of exploring new lines of research.
Some observational facts
concern only a subsample of the stars (for instance the C-rich stars, or the stars
in globular clusters). In those cases 
some special circumstances can be invoked in order to
reproduce the observed features. Other observational facts as the C/O upturn or the
primary nitrogen production point toward some general characteristics shared by a great part if not all the extremely metal poor stars.

\section{What is special in the evolution of extremely metal poor stars ?}

\begin{figure}
 \centering
 \includegraphics[width=9cm, angle=-90]{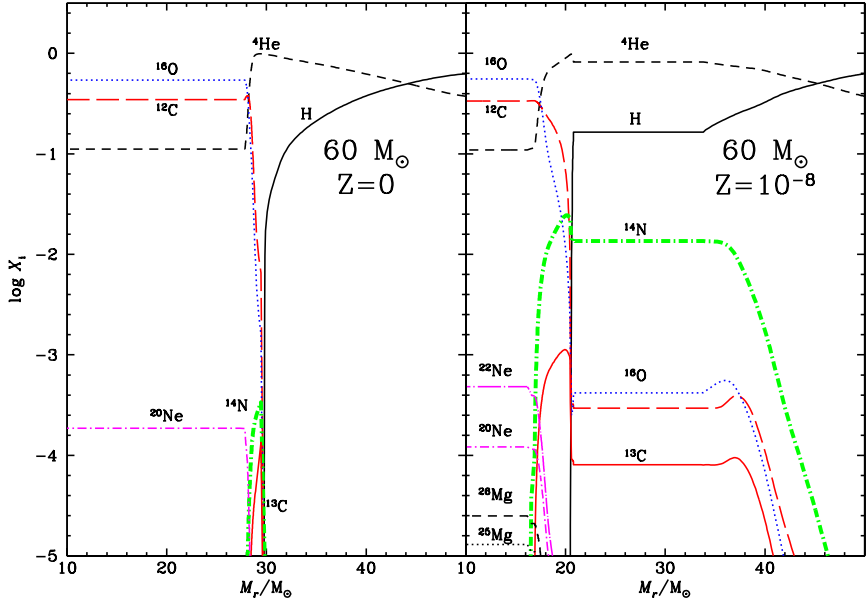}
  \caption{Variations of the abundances (in mass fraction) as a function of the Lagrangian mass
within 60~M$_\odot$ stellar models with
$\upsilon_{\rm ini}$=~800~km~s$^{-1}$ and $Z=0$ and $10^{-8}$, when
the mass fraction of the helium at the center is near 0.1.}\label{fig1}
\end{figure}

Classically much more massive stars can be formed at very low metallicity as a result
of the absence of dust which acts as an efficient cooling agent (see e.g Abel et al.~\cite{Ab04}). The absence of CNO elements modifies the way the H-ignition occurs in massive stars.
The lower opacity of metal poor material 
implies more compact stars and very weak stellar winds. 
Other characteristics might be different in Pop III stars,
for instance the binary frequency, the amplitudes and the effects of magnetic fields. 
Here we want to focus on the effects of rotation. As explained in the contribution
by Maeder et al. in this volume, at low metallicity, the present models predict that
stars have more chance to reach break-up during the Main-Sequence phase, are more efficiently
mixed and may undergo strong mass loss through stellar winds. Here we want to present
some new results illustrating this behavior and draw some consequences for nucleosynthesis
at very low metallicity.

The case of rotating Pop III models is briefly discussed elsewhere in the present volume
(see the contribution by Ekstr\"om et al). In the present paper, we shall concentrate on stars with a very
small initial amount of metals. It is
interesting to point out that already a very small amount of metals makes
a world of difference with respect to strictly Pop III stars. This is illustrated in Fig.~\ref{fig1},
presenting the chemical structure for two rotating 60M$_\odot$ stars 
with $\upsilon_{\rm ini}$ = 800 km s$^{-1}$.
Three differences between the Pop III and the $Z=10^{-8}$ model can be seen: In the $Z=10^{-8}$ stellar model 1) the CO core is smaller; 2) the quantity of primary nitrogen is much greater; 3)
an extended convective zone is associated to the H-burning shell (from about 20 to 34 M$_\odot$).

In the  $Z=10^{-8}$
stellar model, the CNO content in the region of the star where the H-shell ignites 
is already sufficient for the H-burning
to occur through the CNO cycle. This implies that from its birth the H-burning shell can compensate
for a great part of the energy lost by the surface. Typically, the
H-shell luminosity is about one half the total one. The star then
adjusts its structure so that the He-burning core has only to compensate for the other half
of the total luminosity. Instead in the Pop III model, the CNO content in the region
where the H-burning shell occurs is so low that the shell only succeeds in compensating
for a very small fraction of the total luminosity (typically a few percents), the rest having
to be compensated by the He-burning core. This explains the much bigger helium core in the Pop III model
and the very modest H-burning shell. As a consequence in the Pop III model, much less primary nitrogen
is produced and the H-shell remains radiative for a much longer period.


\section{Near break-up limit evolution during the Main-Sequence phase}

\begin{figure}
 \centering
 \includegraphics[width=8.5cm, angle=-90]{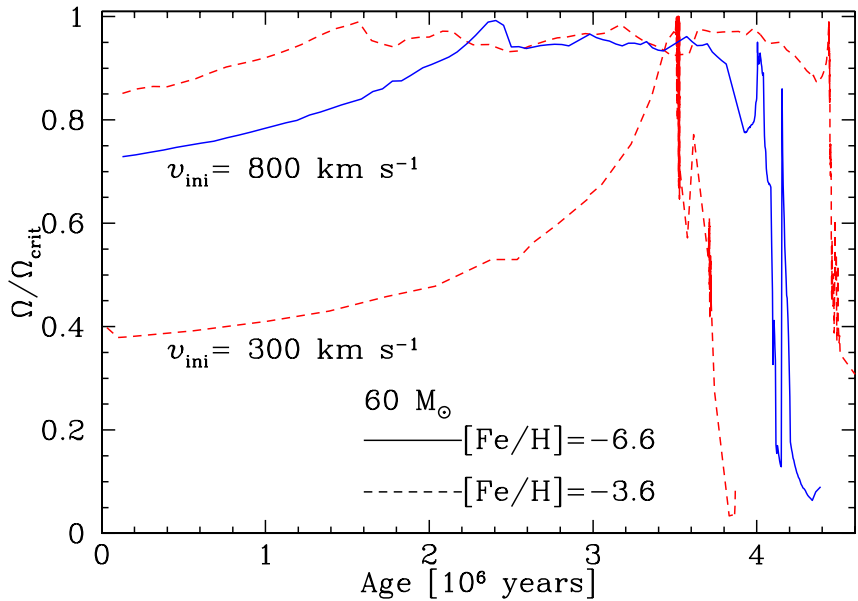}
  \caption{Evolution of $\Omega/\Omega_{\rm crit}$ at the surface of 60~M$_\odot$ models at
$Z=10^{-8}$ (continuous line) and $Z=10^{-5}$ (upper dashed line) with $\upsilon_{\rm
ini}$=~800 km s$^{-1}$. The case of the 60 M$_\odot$ model
at $Z=10^{-5}$ with $\upsilon_{\rm ini}$=~300 km s$^{-1}$
from Meynet \& Maeder~(\cite{MMVIII}) is also shown (lower dashed line).}\label{fig2}
\end{figure}

We do not know what was the initial rotation of the massive first stellar generations.
On the other hand, we can observe the rotational velocity of solar metallicity
massive stars, and suppose that the very metal poor stars had, at the beginning of their evolution,
about the same angular momentum as their solar metallicity counterparts.
For Pop III stars this implies an initial velocity of about 800 km s$^{-1}$.
The same initial velocity of 800 km s$^{-1}$ was considered for the very small metallicities 
$Z=10^{-8}$ and $10^{-5}$ considered in the present paper.

The evolution we shall describe is more dependent on the choice
of the initial velocity than on the precise choice of the initial metallicity. Both
rotating models at $Z=10^{-8}$ and $10^{-5}$ present the same qualitative behaviour, namely they
both produce large amounts of primary nitrogen and lose a great part of their initial mass through
stellar winds.

Fig.~\ref{fig2} shows the evolution of the ratio $\Omega/\Omega_{\rm crit}$ 
at the surface during the MS phase. At both $Z=10^{-8}$ and $10^{-5}$, the fast models
reach the break-up limit, the ``metal-rich'' model at an earlier stage than the
``metal-poor'' one. This comes from the fact that when the metallicity increases, a given value of the initial 
velocity corresponds to a higher initial value
of the $\upsilon_{\rm ini}/\upsilon_{\rm crit}$ ratio.
The stars then remain at
break-up for the rest of their MS lifetimes and undergo an enhancement of their mass loss rates.
As a consequence, the $Z=10^{-8}$ model ends its MS life with 57.6~M$_\odot$, having lost 4\% of
its initial mass, while the $Z=10^{-5}$ model ends its MS life with 53.8~M$_\odot$, having lost 10\% of
its initial mass. 

Despite the stars stay in the vicinity of the break-up limit during an important part of
their MS lifetime, they do not lose very important amounts of mass. 
This is due to the fact that only the outermost layers of 
the stars are above the break--up limit and are ejected. These layers have low density and thus contain little mass.

The 60~M$_\odot$ model at $Z=~10^{-5}$ with
$\upsilon_{\rm ini}$=~300 km s$^{-1}$ computed by Meynet \& Maeder~(\cite{MMVIII})
reaches the break-up velocity much later, only at the end of the MS phase. 
At $Z=10^{-5}$, the velocity 300 km s$^{-1}$ appears thus as the lower limit
for the initial rotation,
allowing a 60 M$_\odot$ star to reach the break-up limit during its MS phase.
This 60 M$_\odot$ star ends its MS life with 59.7~M$_\odot$, having lost only 0.5\% of its
initial mass.

During the MS phase, the surface of the rotating stars is 
enriched in nitrogen and {\bf depleted in carbon} as a result of rotational mixing.
The N/C ratios are enhanced by more than two orders of magnitude 
at the end of the H-burning phase. The total amount of CNO elements
remains however constant.

\begin{figure}
\centering
 \includegraphics[width=8.0cm, angle=-90]{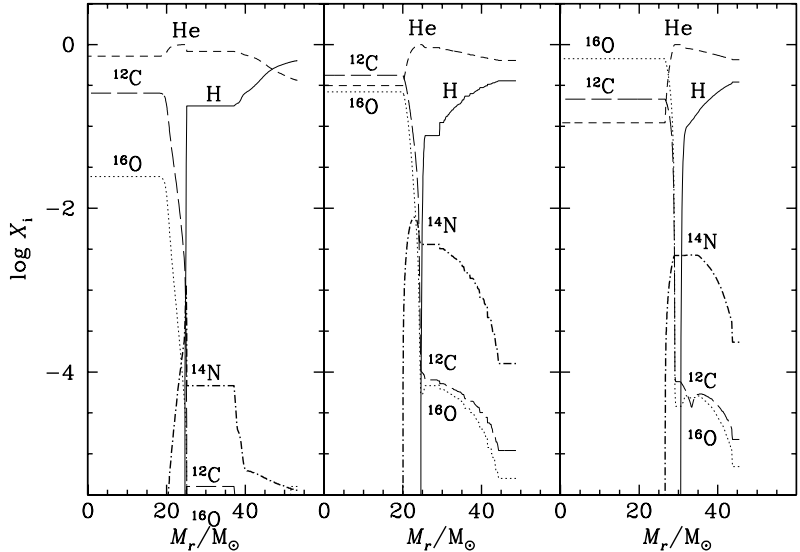} 
  \caption{Chemical composition of a 60 M$_\odot$ stellar model at $Z=10^{-5}$ with
$\upsilon_{\rm ini}$= 800 km s$^{-1}$ when it evolves from the blue to the red
part of the HR diagram. The model shown in the left panel has
$\log L/{\rm L}_\odot = 6.129$ and $\log T_{\rm eff}=4.243$; in the middle
panel, it has $\log L/{\rm L}_\odot = 6.130$ and $\log T_{\rm eff}=4.047$; in the
right panel, it has $\log L/{\rm L}_\odot = 6.145$ and $\log T_{\rm eff}=3.853$.}\label{fig4}
\end{figure}

\section{The post Main-Sequence evolution}

During the core He-burning phase, primary nitrogen is synthesized in the H-burning shell, due
to the rotational diffusion of carbon and oxygen produced in the helium core into the H-burning shell 
(Meynet \& Maeder \cite{MMVIII}). This is well illustrated in 
Fig.~\ref{fig4} for the $Z=10^{-5}$ rotating model. 
 
In contrast to what happens during the MS phase,
rotational mixing, during the core He-burning phase,
induces strong changes of the surface metallicity.
These changes only occur at the end of the core He-burning phase,
when an outer convective zone appears
and rapidly deepens in mass,
dredging up newly synthesized elements to the surface.
From this stage onwards
the surface metallicity increases in a spectacular way.
Typically our 60 M$_\odot$ model at $Z=10^{-8}$ ends its life
with a surface metallicity one million times higher than its initial metallicity, {\it i.e.} with
$Z=0.010$, a metallicity greater that the metallicity
of the Large Magellanic Cloud ! This increase of the surface metallicity
is entirely due to the arrival in great quantities at the surface of primary
CNO elements, carbon and oxygen being produced in the He-core and the nitrogen
in the H-shell.

The consequence of such important surface enrichments 
on the mass loss rates remains to be studied in details 
using models of stellar winds with the appropriate physical 
characteristics (position in the HR diagram and chemical composition). 
In absence of such sophisticated models, we applied here 
the usual rule, namely $\dot M(Z)=(Z/Z_\odot)^{1/2}\dot M(Z_\odot)$, where $Z$
is the metallicity of the outer layers. With this prescription, the stars
lose a great part of its initial mass. Our 60 M$_\odot$ at $Z=10^{-8}$
ends its life with only 24 M$_\odot$, the corresponding model at $Z=10^{-5}$
reaches a final mass of 37 M$_\odot$.

\begin{figure}
 \centering
 \includegraphics[width=9.0cm, angle=-90]{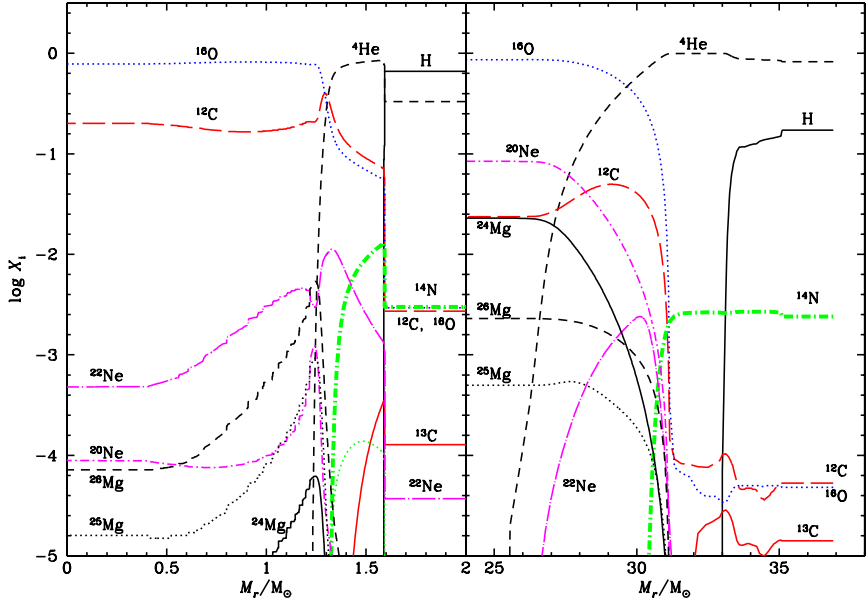}
  \caption{Variations of the abundances (in mass fraction) as a function of the Lagrangian mass
within a 7~M$_\odot$ stellar model in the early AGB phase, and at the end of the core
C-burning phase in a 60~M$_\odot$ stellar model. Both models were computed
with $\upsilon_{\rm ini}$=~800~km~s$^{-1}$. The metallicity is $Z=10^{-5}$.}\label{fig5}
\end{figure}

We have also computed a 7 M$_\odot$ stellar model at $Z=10^{-5}$ with
$\upsilon_{\rm ini}= 800$ km s$^{-1}$. This model, in contrast with its more
massive counterpart, never reaches the break-up limit during the MS phase.
The 7 M$_\odot$ is more compact than the 60M$_\odot$ model, this makes
the Gratton-\"Opick term in the expression for the velocity of the meridional
circulation smaller (see the contribution by Maeder et al. in this volume). The angular
momentum is thus less efficiently transported outwards than in the more massive model,
making the reaching of the break-up limit more difficult. Interestingly, the model
present a higher degree of mixing than the 60 M$_\odot$ stellar model, because,
due to the relative inefficiency of the angular momentum transport, the
gradients of $\Omega$ remain steeper, making the shear diffusion stronger. 
This is in contrast to what happens at higher metallicity.
Indeed it was shown (see Maeder \& Meynet~\cite{MMVII}) that at a given metallicity and
for a given initial velocity, the mixing
was more efficient in the more massive stars. This is correct as long as the gradients
of $\Omega$ do not depend too much on the initial mass. Here at very low $Z$,
the gradients are very sensitive to the initial mass, being steeper in smaller initial mass stars.

The 7 M$_\odot$ remains in the blue part of the HR diagram during the whole He-burning phase,
preventing an outer convective zone to appear and to dredge-up the primary CNO elements.
Only at the end of the core He-burning phase, the star evolves to the red and approach
the base of the Asymptotic Giant Branch. At this point, an outer convective zone appears
and produces an enormous enhancement of the surface metallicity. This can be seen
in Fig.~\ref{fig5}, which compares the chemical structure of the 7 M$_\odot$ at the
early AGB phase, with that of the 
60 M$_\odot$ model at the end of the core C-burning phase. We see that in the outer layer
of the 7 M$_\odot$ the abundances of CNO elements are at about the same level, while
in the 60 M$_\odot$, the abundance of nitrogen is about at the same level as in the
7 M$_\odot$ model, while those of carbon and oxygen are many orders of magnitude below. 

\begin{figure}
 \centering
 \includegraphics[width=10.0cm, angle=0]{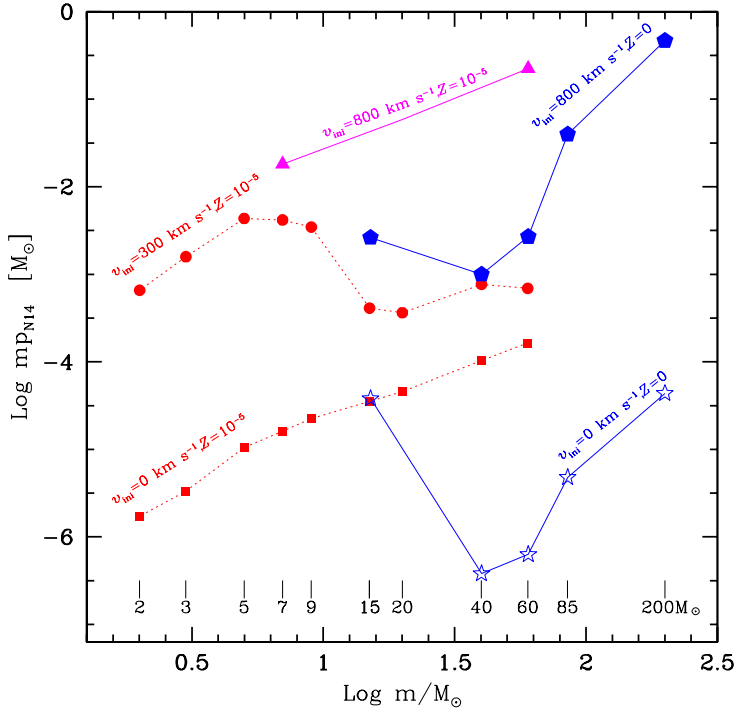}
  \caption{Stellar yields in N from different initial mass stars 
at $Z=0$, $10^{-8}$ and $10^{-5}$ with and without rotation. 
The contribution of the stellar winds have been accounted for when present.}\label{fig7}
\end{figure}

\section{Consequences for the CNO yields}

Let us first discuss the CNO yields of Pop III stellar models (see Ekstr\"om et al. in this volume). 
The yields of carbon and oxygen appear to be little affected by
rotation, while those of nitrogen are enhanced by 3 to 4 orders of magnitude in rotating models.
The yields for nitrogen are compared with those obtained in other models in Fig.~\ref{fig7}.
We see that the Pop III yields from rotating model with $\upsilon_{\rm ini}=800$ km s$^{-1}$
are in general higher than those for $Z=10^{-5}$ with $\upsilon_{\rm ini}=300$ km s$^{-1}$
computed by Meynet \& Maeder~(\cite{MMVIII}). However, according to Chiappini et al.~(\cite{Ch05}),
the yields from massive stars, required to fit the N/O ratios in the unmixed sample of halo stars
observed by Israelian et al.~(\cite{Is04}) and
Spite et al. (\cite{Sp05}), should be enhanced by a factor 50 (1.7 dex), {\it i.e.} should be much higher
than predicted by the present Pop III stars for stars in the mass range between 15 and 85 M$_\odot$.
Unless, only very massive stars are formed, it appears thus difficult to explain the high
N/O ratios observed in halo stars. Moreover it is likely that the production of primary nitrogen
by massive stars extends over a range of metallicities and is not restricted to Pop III stars.
As shown above, a tiny amount of metal can boost
in a very important way the amount of primary nitrogen produced. In  that case, starting
with a velocity of $\upsilon_{\rm ini}=800$ km s$^{-1}$ would allow to reach
levels well above those required by Chiappini et al.~(\cite{Ch05}) to fit the
observed N/O in halo stars.
Thus rotating massive star models have no difficulty in making the great amounts
of primary nitrogen which seem to be required by the observations.
More work is still needed in order to explore the range of initial
conditions which would give the best agreement.

\section{New scenarios for the origin of the C-rich stars}

\begin{figure}
 \centering
 \includegraphics[width=7.0cm, angle=-90]{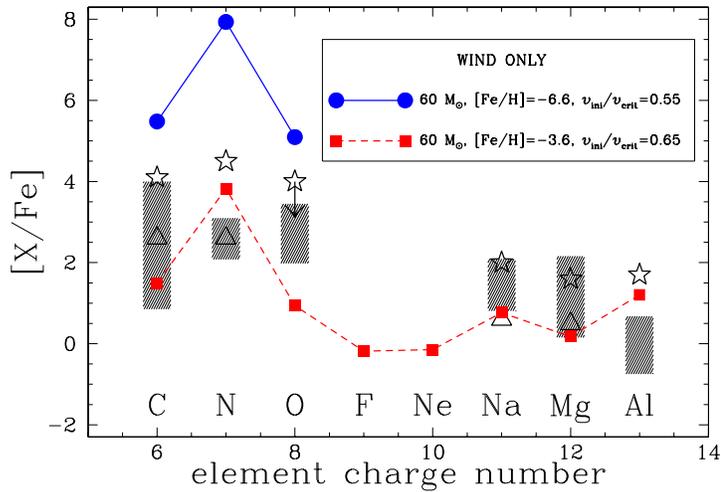}
  \caption{Chemical composition of the wind
of rotating 60~M$_\odot$ models (solid circles and squares).
The hatched areas correspond to the range of values
measured at the surface of giant CEMP stars: HE 0107-5240, [Fe/H]$\simeq$~-5.3 
(Christlieb \& al. \cite{christ04});
CS 22949-037, [Fe/H]$\simeq$~-4.0 (Norris \& al.
\cite{norr01}; Depagne \& al. \cite{dep02}); CS 29498-043, [Fe/H]$\simeq$~-3.5 (Aoki
\& al. \cite{aoki04}). The empty triangles (Plez \& Cohen~\cite{Pl05}, [Fe/H]$\simeq -4.0$) 
and stars (Frebel et al.~\cite{Fr05},  [Fe/H]$\simeq -5.4$, only an upper limit is given for [O/Fe]) corresponding to
non-evolved CEMP stars (see text).}\label{fig8}
\end{figure}

\begin{figure}
\centering
 \includegraphics[width=7.0cm, angle=-90]{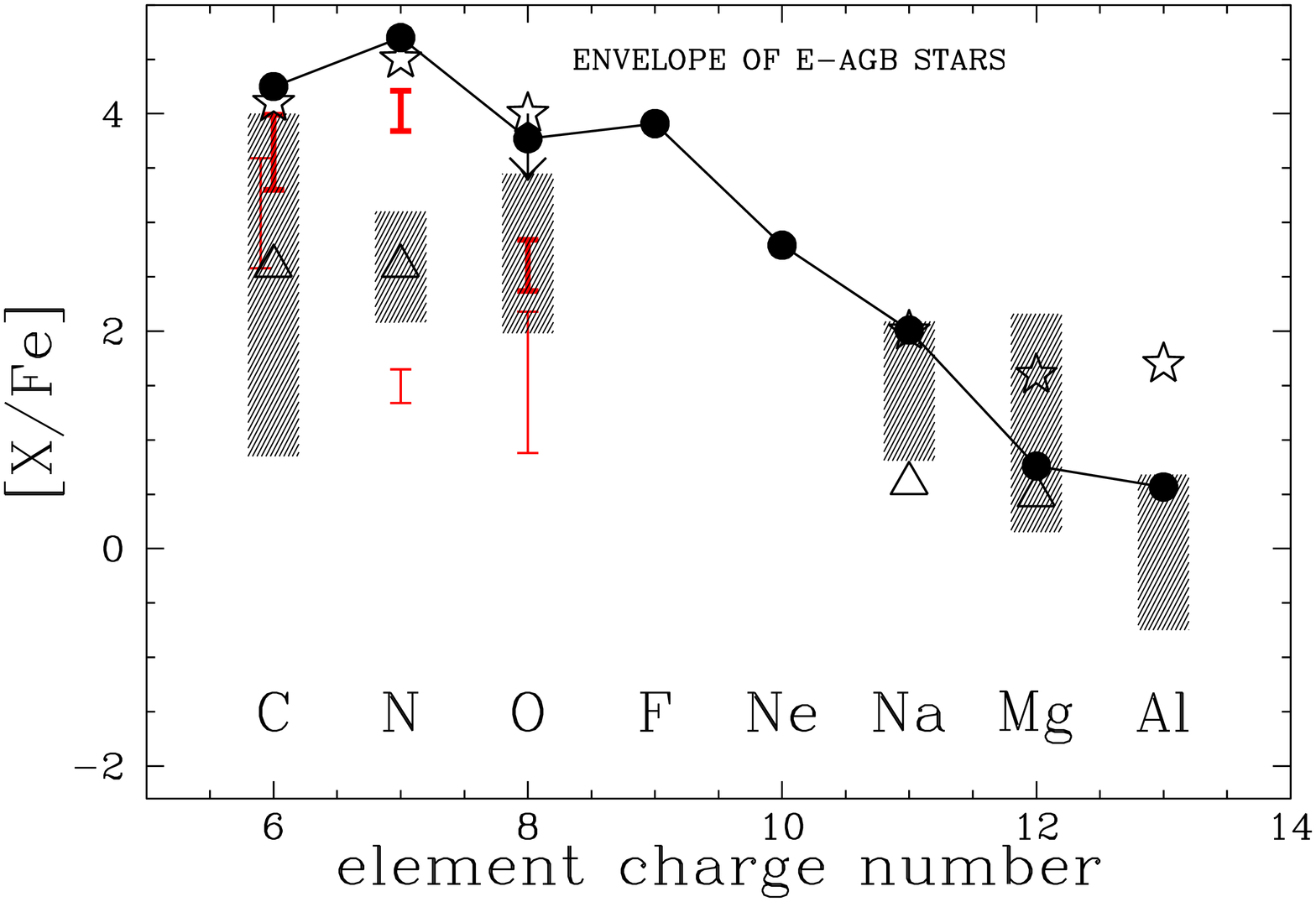}
  \caption{Chemical composition of envelopes of E-AGB stars compared to abundances
observed at the surface of CEMP stars (hatched areas, same observations
as in Fig.~\ref{fig8}). The continuous line shows the case
of a 7 M$_\odot$
at $Z=10^{-5}$ ([Fe/H]=-3.6) with $\upsilon_{\rm ini}=$ 800 km s$^{-1}$.
The vertical lines (shown as ``error bars'')
indicate the 
ranges of values for CNO elements
for the stellar models from Meynet \& Maeder~(\cite{MMVIII})
(models with initial masses between 2 and 7 M$_\odot$ at $Z=10^{-5}$).
The thick and thin lines correspond to rotating ($\upsilon_{\rm ini}$ = 300 km s$^{-1}$)
and non-rotating models.}
  \label{fig9}
\end{figure}

Spectroscopic surveys of very metal poor stars (Beers et al.~\cite{Be92}; 
Beers~\cite{Be99}; Christ\-lieb~\cite{Ch03}) 
have shown that Carbon-rich Extremely Metal Poor stars (CEMP stars)
account for up to about 25\% of stars with metallicities lower than 
[Fe/H]$\sim -2.5$.
A star is said to be C-rich if [C/Fe]$>1$.
The two most iron-deficient stars observed so far, 
HE 0107-5240, a giant halo star, and HE 1327-2326, a dwarf or subgiant halo star,
are C-rich stars. This might indicate that
the frequency of C-rich stars increases when the metallicity decreases.
Observations of C-rich non-evolved stars ({\it i.e.} dwarf or subgiant) indicate also
that the pattern of abundances was already present in the cloud from which the star formed
and is therefore not due to a mechanism occurring in the star itself.

Many scenarios have already been proposed to explain the very peculiar abundances
at the surface of these stars (supernova with mixing and strong fall back, Umeda \& Nomoto~\cite{Um03};
mixing of the ejecta of two supernovae, Limongi et al.~\cite{Li03}; mass transfer from an AGB star, Suda
et al.~\cite{Su04}). Here we propose new scenarios based on rotating models.

Let us first see if the CEMP stars could be formed from material made up of massive star wind
(or at least heavily enriched by winds of massive stars). At first sight such
a model might appear quite unrealistic, since the period of strong stellar winds
is rapidly followed by the supernova ejection, which would add to the wind ejecta
the ejecta of the supernova itself. However,
for massive stars, it might occur that at the end of their nuclear lifetime, a black
hole, swallowing the whole final mass, is produced. In that case, the massive star
would contribute to the local chemical enrichment of the interstellar medium only through
its winds. Let us suppose that such a situation has occurred and that
the small halo star that we observe today 
formed from the shock induced by the stellar winds with the interstellar material.
What would be its chemical composition ? 
Its iron content would be the same as that
of the massive star
since the iron abundance in the interstellar medium had no
time to change much in the brief massive star lifetime. 
Also, the massive star wind ejecta 
are neither depleted nor enriched in iron.
The abundances of the other elements in the stellar winds for our two rotating 60 M$_\odot$
at $Z=10^{-8}$ and 10$^{-5}$ are shown in Fig.~\ref{fig8}.

For the two metallicities considered here,
the wind material of rotating models is characterized by N/C and N/O ratios 
between
$\sim$ 1 and 40, and $^{12}$C/$^{13}$ C ratios around 4-5. These values are
compatible with the ratios observed at
the surface of CS 22949-037 (Depagne et al.~\cite{dep02}):
N/C $\sim$ 3 and $^{12}$C/$^{13}$C $\sim$4. The observed value
for N/O ($\sim$0.2) is smaller than the range of theoretical values, but greater
than the solar ratio ($\sim$ 0.03). Thus the observed N/O ratio
also bears the mark of some CNO processing, although slightly less
developed than in our stellar wind models. On the whole, 
a stellar wind origin for the material composing this star
does not appear out of order in view of the above comparisons,
especially
if one considers the fact that, in the present comparison, there is no
fine tuning of some parameters in order to obtain the best agreement possible. 
The theoretical results are
directly compared to the observations. Moreover only a small subset of possible initial conditions
has been explored.

Other CEMP stars present however lower values for the N/C and N/O ratios
and higher values for the $^{12}$C/$^{13}$C ratio.
For these cases it appears that the winds of our rotating 60 M$_\odot$ models
appear to be too strongly CNO processed (too
high N/C and N/O ratios and too low $^{12}$C/$^{13}$C ratios).
Better agreement would
be obtained if the observed abundances also result from material from the CO-core, ejected
either by strong late stellar winds or in supernova explosion.

To explore this possibility we have study the case where wind ejecta are mixed with
supernova ejecta and interstellar material. At least
two free parameters have to be introduced: 1) the dilution
factor between the ejecta and the interstellar medium; 2) the mass of iron in the
supernova ejecta. It is possible to adopt values of these two parameters so that the
observed values of [Fe/H], [O/Fe] and [C/Fe] can be reproduced by both our non-rotating
and rotating 60 M$_\odot$ model. However, the observed [N/Fe] ratios can only be obtained
using the rotating model. The wind and supernova ejecta model lowers the N/C and N/O ratios
improving the agreement with respect to the pure wind model, however at the cost of 
two free parameters ! 

Another model has been suggested by Suda et al.~(\cite{Su04}): the small halo star,
observed today, was the secondary in a binary system whose primary went through
the AGB stage. At this stage, part of the AGB envelope has been 
accreted by the secondary greatly modifying its original surface abundances.
As was the case for the massive star wind model seen above , the iron must come from a previous
star generation, since no iron is produced by the AGB star.

Using our models computed in Meynet \& Maeder~(\cite{MMVIII}) for 
initial masses between 2 and 7 M$_\odot$ at $Z=10^{-5}$
and
with $\upsilon_{\rm ini}=0$ and $300$ km s$^{-1}$,
we can estimate the chemical composition 
of the envelope of intermediate mass stars 
at the beginning of the thermal pulse Asymptotic Giant Branch phase. 
The envelope is all the matter above the CO-core.
The range of values for the CNO elements given by these models are shown in Fig.~\ref{fig9}
(vertical thin lines: no-rotation; vertical thick lines, with rotation).
We have also computed a new
7 M$_\odot$ with
$\upsilon_{\rm ini}=800$ km s$^{-1}$ at $Z=10^{-5}$ (continuous
line with solid circles). 

We see that the envelopes of AGB stellar models with rotation 
show a chemical composition very similar to that observed at the surface
of CEMP stars. In particular, only rotating models are able to simultaneously explain
the large abundances of C, N and O.
It is however difficult to conclude that rotating intermediate mass star models
are better than rotating massive star models in reproducing the
abundance pattern of CEMP stars. Probably,
some CEMP stars are formed from massive star ejecta and others
from AGB star envelopes. Interestingly at this stage,
some possible ways to distinguish between massive star wind material
and AGB envelope do appear. Indeed,
massive star wind material is characterized by very low $^{12}$C/$^{13}$C ratio,
while intermediate mass stars seem to present higher values for this ratio.
AGB envelopes would also present very high overabundances
of $^{17}$O, $^{18}$O, $^{19}$F and $^{22}$Ne, while wind of massive rotating
stars present a weaker overabundance of $^{17}$O and depletion of 
$^{18}$O, $^{19}$F and $^{22}$Ne.
As discussed in Frebel et al.~(\cite{Fr05}), the ratio of heavy elements
as the strontium to barium ratio can also give clues as to the origin of the material
from which the star formed. In the case of HE 1327-2326, Frebel et al.~(\cite{Fr05})
give a lower limit of [Sr/Ba] $> -0.4$, which suggests that strontium was not
produced in the main s-process occurring in AGB stars, leaving thus
the massive star hypothesis as the best option, in agreement with the result
from $^{12}$C/$^{13}$C in G77-61 (Plez \& Cohen~\cite{Pl05}) and
CS 22949-037 (Depagne et al.~\cite{dep02}).



\end{document}